# Extraordinary Sound Transmission through Density-Near-Zero Ultranarrow Channels

Romain Fleury, Andrea Alù[*]

Dept. of Electrical & Computer Engineering, The University of Texas at Austin

*alu@mail.utexas.edu

*We introduce the acoustic equivalent of 'supercoupling' by studying the anomalous sound transmission and uniform energy squeezing through ultranarrow acoustic channels filled with zero-density metamaterials. We propose their realization by inserting transverse membranes with a subwavelength period along the channel, and we prove a novel form of acoustic tunneling based on impedance matching and infinite phase velocity at the zero-density operation. We envision applications in sensing, noise control, cloaking and energy harvesting.*

Artificial materials with anomalous values of constitutive parameters have been extensively studied over the past decades. Electromagnetic (EM) metamaterials with extreme constitutive parameters [1], very low [2-7] or very high [8] compared to those available in nature, are often associated with spectacular phenomena, like anomalous wave tunneling and negative refraction. Zero permittivity (ENZ) metamaterials have been proposed for a variety of exciting applications, including cloaking [9] and radiation patterning [2]. In these materials, waves propagate with very large phase velocity, effectively producing quasi-static field distributions. An exciting application of these properties consists in the anomalous tunneling and matched transmission through very narrow channels filled with ENZ materials, also known as 'supercoupling' [4,10]. This effect is substantially different from conventional Fabry–Pérot (FP) tunneling resonances through mismatched channels or apertures, as it is based on impedance matching between waveguides with very different geometric cross-section, which is compensated by the anomalously large wave impedance in ENZ materials [4]. This anomalous tunneling is accompanied by *large* and *uniform* field enhancement along the channel, due to the quasi-static nature of wave propagation in ENZ, implying that matching and tunneling do not depend on the channel length or on the presence of bends and twists along the channel. These properties make supercoupling an ideal phenomenon for coupling distant waveguides and antennas with good matching and no phase delay [6], for filtering independent of the channel geometry [4], light concentration and harvesting [11], sensing [12], boosting molecular emission [13] and nonlinearities [14-15]. Translating these concepts to acoustic waves may have several additional advantages: nonlinearities are inherently stronger for acoustic waves and guided acoustic waves have different dispersion properties that may be tailored to our advantage, as we outline in the following. Here we highlight analogies, differences and challenges associated with acoustic impedance matching and anomalous sound tunneling through ultranarrow channels for acoustic waves. As in the EM scenario, we rely on the extreme value of one of the constitutive parameters in an ultranarrow tube to *totally compensate* the geometrical mismatch at the connection with a much larger feeding waveguide. At the tunneling frequency, acoustic energy can be concentrated within a very small transverse cross-section, leading to interesting possibilities in a variety of acoustic applications.

Consider the geometry sketched in the inset of Fig. 1: two identical circular cylindrical acoustic waveguides with cross-sectional area $S_{wg}$ are connected through a much thinner tube with $S_{ch} \ll S_{wg}$. Assuming that the waveguides support the dominant transverse mode, transmission-line theory may be used to calculate the reflection and transmission coefficients for such a connection. In this analytical model, the line voltage is defined as the acoustic pressure $p$ while the line current is associated with the total volumetric flow $q_z$ through the waveguide cross-section. These definitions ensure the continuity of pressure and volumetric flow at a junction between two lines [16,17]. The reflection coefficient at the channel entrance may be generally written as

$$R = \frac{(Z_{ch}^2 - Z_{wg}^2)\tan(\beta_{ch}l)}{(Z_{ch}^2 + Z_{wg}^2)\tan(\beta_{ch}l) + 2iZ_{ch}Z_{wg}}, \qquad (1)$$

where $Z_{wg}$ and $Z_{ch}$ are the line impedances associated with the feeding waveguide and the small channel respectively, $l$ is the channel length and $\beta_{ch}$ is the wave number in the channel under an $e^{-i\omega t}$ time convention. Total transmission is obtained at the usual FP resonances when $\beta_{ch}l = n\pi$, with $n$ being an integer. Another possibility for zero reflection arises under the condition $Z_{ch} = Z_{wg}$, i.e. when the line impedances are *matched*. This matching condition may be explicitly written in terms of the normalized acoustic impedance of the two waveguides as:

$$\frac{\sqrt{\rho_{ch}\kappa_{ch}}}{S_{ch}} = \frac{\sqrt{\rho_{wg}\kappa_{wg}}}{S_{wg}}, \qquad (2)$$

where $\rho_{ch}, \rho_{wg}$ are the effective densities of the acoustic materials filling the waveguides and $\kappa_{ch}, \kappa_{wg}$ are the corresponding bulk moduli. Note that the expression of line impedance differs from the characteristic impedance of the acoustic medium by a factor $1/S$. This is required by the continuity of the volumetric flow at the abruptions.

If the big waveguides are filled by a conventional material, like air, the matching condition (2) can be met as long as the characteristic impedance of the channel $\sqrt{\rho_{ch}\kappa_{ch}}$ takes extremely low values. In the limit $\sqrt{\rho_{ch}\kappa_{ch}} \to 0$, total transmission may be counterintuitively achieved for a channel of infinitesimal cross-section. This extreme value of *characteristic* impedance can exactly compensate the mismatch introduced in the *line* impedances by the difference in cross sections. This is the dual of the extremely *large* electric impedance required in the analogous EM problem, which is usually achieved with ENZ metamaterials.

There are several ways to modify the effective acoustic parameters to bring the impedance to a very low value [18-21]. If we focus on modifying only one parameter (effective density or bulk modulus), the wave number $\beta_{ch} = \sqrt{\rho_{ch}\kappa_{ch}^{-1}}$ is also required to take an extreme value with anomalous propagation properties: if $\rho_{ch} \to 0$ (DNZ), total transmission will be associated with uniform phase along the channel, or *very fast* waves. This is the acoustic equivalent of ENZ supercoupling. If $\kappa_{ch} \to 0$ (KNZ), impedance matching and total transmission will be associated with very large wave numbers and *very slow* waves. This situation would be equivalent to EM supercoupling based on large permeability $\mu$ (MVL), a scenario not considered in previous works.

Another remarkable difference between EM and acoustic scenarios consists in the modal propagation and dispersion properties. For EM waves, closed waveguides are inherently dispersive, and therefore it is not possible to achieve supercoupling for a waveguide with arbitrarily small cross-section. All theoretical and experimental works on EM supercoupling have been based either on 2-D problems, in which the waveguides are infinite in the transverse direction, or by considering 3-D waveguides with one dimension (the width) comparable to the wavelength of excitation, used to tailor the dispersion and operate the channel near cut-off, which is equivalent to an effective permittivity near zero [23]. In the acoustic case, however, closed connected 3-D waveguides do support longitudinal modes with no frequency dispersion and no cut-off, which allows us to extend the supercoupling phenomenon truly in 3-D. The price to be paid is that we do need to realize a metamaterial inside the channel to achieve the required extreme bulk parameters, and we cannot use to our advantage the modal dispersion of the channel to operate near cut-off.

We consider here periodically loading the channel with subwavelength inclusions near resonance, which may realize an acoustic metamaterial with effective density or bulk modulus near-zero [18,19]. We can model the propagation of the dominant mode in a periodically loaded acoustic waveguide again using a transmission-line model, with unit cell schematically shown in the inset of Fig. 1. The bulk modulus $\kappa_0$ and the density $\rho_0$ of the acoustic medium filling the waveguide are associated to the parallel acoustic compliance $C_0$ and the series acoustic mass $m_0$, respectively:

$$\begin{aligned} C_0 &= S\kappa_0^{-1}dz \\ m_0 &= S^{-1}\rho_0 dz \end{aligned} \quad (3)$$

Series loads affect the effective $\rho$ experienced by the mode by modifying the mass (as in the inset of Fig. 1), whereas parallel or shunt loads affect the bulk modulus $\kappa$, modifying its compliance. Consider in this regard a basic subwavelength load consisting of a resonator modeled as a lumped inertance $m_{inc}$ in series with a compliance $C_{inc}$. This may model a subwavelength resonant element loading the channel, like a membrane or a Helmholtz resonator. If such inclusions periodically load the waveguide with a period $d$ small compared to the wavelength, proper averaging may be performed to define effective acoustic parameters. When the inclusions are arranged in a *series* connection, as in the case of a thin membrane clamped to the waveguide walls [16,17,22], the corresponding circuit model is shown in the inset of Fig. 1, and the effective density is modified into

$$\rho_{eff} = \rho_0 + \frac{m_{inc} S}{d}\left(1 - \frac{1}{m_{inc} C_{inc} \omega^2}\right). \tag{4}$$

Similarly, for a *parallel* load as in the case of a Helmholtz resonator connected to one of the channel walls [20], the effective bulk modulus becomes:

$$\frac{1}{\kappa_{eff}} = \frac{1}{\kappa_0} + \frac{C_{inc}}{dS}\frac{1}{1 - m_{inc} C_{inc} \omega^2}. \tag{5}$$

Eq. (4) and (5) show that a simple resonant series $m_{inc}$-$C_{inc}$ load can tune the effective parameters of the waveguide, allowing density or bulk modulus to assume negative, zero or positive values, depending on the frequency and type of connection.

For anomalous matching and sound tunneling, we look at the conditions for which $\rho_{eff}$ or $\kappa_{eff}$ are near zero, which arise close to the inclusion resonance [slightly shifted in the density case, according to Eq. (4)]. In the zero-modulus case, however, $\kappa_{eff}$ cannot be made arbitrarily small since, when it approaches $0^+$, the wave number $\beta_{ch}$ increases and the corresponding wavelength $\lambda_{ch} = 2\pi / \beta_{ch}$ decreases becoming comparable to the period $d$. In this limit, the averaging (5) is no longer applicable. In particular, for $\beta_{ch} > \pi / d$ or $\kappa_{eff} < \rho_{ch} d^2 / \pi^2$ the periodically loaded channel hits a bandgap and the dominant mode cannot propagate. Since we cannot get too small values of $\kappa_{eff}$, according to (2) there is a lower limit to the cross-sectional mismatch achievable to be able to observe matched transmission in the KNZ case. Because the wavelength shrinks for smaller values of $\kappa_{eff}$, we also expect a large number of tunneling resonances around the KNZ tunneling peak, which decreases the overall bandwidth. On the contrary, in the DNZ case the wave number gets larger for smaller densities, creating a quasi-static field distribution that reinforces the averaging requirements to derive (4). We conclude that, although in principle viable, the KNZ case is less interesting for applications involving a realistic periodic metamaterial compared to the dual DNZ case, and in the following we focus on this DNZ tunneling condition.

We validate this simple homogenization model in the DNZ limit comparing our analytical model with full-wave simulations. In our analysis, we model the membranes as punctual acoustic series impedances $Z_{inc}$, periodically loading an air-filled waveguide with line impedance $Z_{0,ch} = \sqrt{\rho_0 \kappa_0} S_{ch}^{-1}$ and propagation constant $\beta_0 = \omega c_0^{-1}$. The channel is connected at its ends to two semi-infinite waveguides with line impedance $Z_{0,wg} = \sqrt{\rho_0 \kappa_0} S_{wg}^{-1}$. The reflection coefficient $R$ at the entrance of the channel may be calculated as

$$R = \frac{Z - Z_{0,wg}}{Z + Z_{0,wg}} \qquad (6)$$

where Z is the input impedance at the entrance of the channel, calculated recursively using

$$\eta_{i+1} = \left( Z_{0,ch} \frac{\eta_i - iZ_{0,ch} \tan(\beta_0 d)}{Z_{0,ch} - i\eta_i \tan(\beta_0 d)} \right) + Z_{inc} \qquad (7)$$

with

$$\eta_1 = \left( Z_{0,ch} \frac{Z_{0,wg} - iZ_{0,ch} \tan(\beta_0 \frac{d}{2})}{Z_{0,ch} - iZ_{0,wg} \tan(\beta_0 \frac{d}{2})} \right) + Z_{inc} \qquad (8)$$

and

$$Z = Z_{0,ch} \frac{\eta_N - iZ_{0,ch} \tan(\beta_0 \frac{d}{2})}{Z_{0,ch} - i\eta_N \tan(\beta_0 \frac{d}{2})}. \qquad (9)$$

The integer $N$ in (9) is the total number of unit cells in the channel. The additional reactance at the ends of the channel, associated with higher-order evanescent modes excited at the abruptions and the non-idealities in the membranes, can be accounted for by slightly modifying the effective channel length [17]. Full-wave simulations were performed using Comsol Multiphysics to accurately model the acoustic-structure interaction, with fully coupled structural mechanics and linear acoustic equations. The membranes are considered in their full thickness and modeled as linear elastic solids, while the acoustic medium is assumed to be linear and homogeneous. The waveguide walls are modeled as hard boundaries.

Fig. 2 shows the transmission coefficient versus frequency for two circular waveguides of 25mm radius connected by a channel of 2.5mm radius containing $N$ polyimide membranes of thickness $t = 25$μm and period $d = 10$ mm. In the figure, we show the effect of varying $N = 19$ (blue lines) and $N = 20$ (red lines) on the two resonant peaks in the considered frequency range. Dashed lines correspond to analytical results and solid lines to full-wave simulations. By changing the channel length, the two resonances show a very different response: the first peak occurs at the same frequency, independent of the channel length. This frequency corresponds to the zero density condition derived in (4), for which the channel is impedance *matched* to the waveguides, and tunneling occurs independent of its length. The second peak, on the contrary, shifts towards higher frequencies for shorter channels, as expected for a conventional FP resonance. The curves are well captured by our analytical model, with small discrepancies attributed to the energy stored in the higher-order modes at the junctions and around the membranes, which do not behave like ideal pistons. It is evident that the DNZ peak performs analogously to ENZ supercoupling for EM waves, providing total matched transmission independent of the coupling distance. A clear advantage of this acoustic problem, compared to the EM scenario, is that we are able to shrink the channel in both transverse dimensions, with very appealing potential for sound squeezing and nonlinearity enhancement.

Our model does not only predict anomalous tunneling, but also uniform, quasi-static phase distribution along the channel, and an averaged particle velocity enhanced by a factor equal to the ratio of cross sections. We therefore expect *a uniform velocity boosting* all along the channel of ~100 in our example. In Fig. 2, we show the velocity field distribution within the channel normalized to impinging modal velocity at three different frequencies: below the DNZ point (2.4 kHz), for which Eq. (4) predicts a negative density, at the DNZ operation (2.415 kHz), and at the FP resonance frequency (2.43 kHz). The bottom graphs show the velocity magnitude and phase along the center of the channel. For negative density, the mode does not propagate and the velocity is essentially zero. In the DNZ case, on the contrary, the average velocity is uniform along the channel, with an average gain of ~100, as predicted by our model. Crossing the membranes, the velocity is locally enhanced as the membranes are clamped at the edges, but the enhancement is uniform away from them within each unit cell. The phase variation has a linear evolution in the outer waveguide, as expected for a propagating mode without reflections. Inside the channel at the DNZ operation the phase is very flat, and the wave tunnels through the channel while acquiring almost zero phase change. We observe no phase variation over a distance of about 1.5 times the wavelength in the outer waveguides, ensuring a very large phase velocity within the channel. At the FP resonance, on the contrary, the velocity is largely enhanced at the channel entrance and exit, but it is zero at the center, with a strong standing wave pattern and a 180° phase shift, obviously largely dependent on the channel length.

In Fig. 3 we investigate the effect of losses by varying the loss factor $\eta$ of the membrane, defined as the ratio of energy dissipated per radian oscillation to the maximum strain energy. Both tunneling peaks are quite robust to realistic values of losses, as polyimide has $\eta \simeq 10^{-3}$ at the considered frequency [23] (red line). The DNZ peak is slightly more affected than the FP peak, since the mode is more uniformly enhanced, and therefore feels more strongly the level of losses in the membranes. This may be advantageous for efficient energy harvesting and for enhancing nonlinear effects in the channel [15]. The membranes may be replaced by electrical transducers playing the twofold role of metamaterial inclusions and energy harvesters or nonlinear elements, and the uniform enhancement in the channel allows a more efficient process.

We note that the averaged pressure field, unlike the velocity, is not enhanced at DNZ transmission, consistent with the continuity of power flow along the transmission-line in our model. This implies that losses affecting only the compliance in the channel would not affect the DNZ tunneling, in contrast with conventional FP resonances.

The quasi-static nature of wave tunneling in ENZ channels provides robust EM transmission not only for different lengths of a straight channel, but also in the case of arbitrary twists and/or bends, without affecting the tunneling properties or the frequency of operation [4,24]. We expect even more spectacular performance in this acoustic scenario, as the DNZ operation allows us to bend and twist the channels in all directions, without having to worry about affecting its modal dispersion. We investigate this possibility in Fig. 4, which shows the acoustic pressure field distribution for different bending scenarios. Here, we excite at the DNZ frequency and bend the channel in different ways to explore how this affects the tunneling properties. We choose the maximum radius of curvature in each case to be $\lambda/4$, in order to create short and abrupt transitions, which is the worst case scenario for reflection and frequency detuning. We tested the cases of a 90° bent channel (Fig. 4a), of a 180° bent channel (Fig 4b) and of a longer S-shaped channel (Fig. 4c). This last scenario has a total length about 1.6 wavelengths in free-space. In all cases, anomalous matched tunneling is found at the same frequency, independent of the shape, length and twist, with almost no phase variation all along the channel, due to the uniform phase velocity enhancement. These properties are completely unaffected by the channel curvature, as long as the unit cell volume and membrane geometry remain approximately uniform along the channel.

These findings may have important applications in noise control, sensing, cloaking, sound energy harvesting and nonlinearity enhancement. The extraordinary squeezing capabilities of DNZ ultranarrow channels may be exploited to boost nonlinearities for sensing, switches and parametric amplifiers. Ultranarrow DNZ channels, obtained by simply loading a capillary tube with membranes, may be used to drastically change the boundary conditions of an acoustic cavity, with interesting applications in modal noise control of rooms. The membranes could be also replaced by active transducers to create self-tunable energy harvesters or dissipaters. Finally, we believe that the capability of bent DNZ channels to deviate the flow of acoustic energy could be exploited in cloaking applications and in noise control.

This work has been supported by the DTRA YIP award No. HDTRA1-12-1-0022. The authors are grateful to Dr. Michael Haberman and Mr. Caleb Sieck for useful discussions.

**Figures**

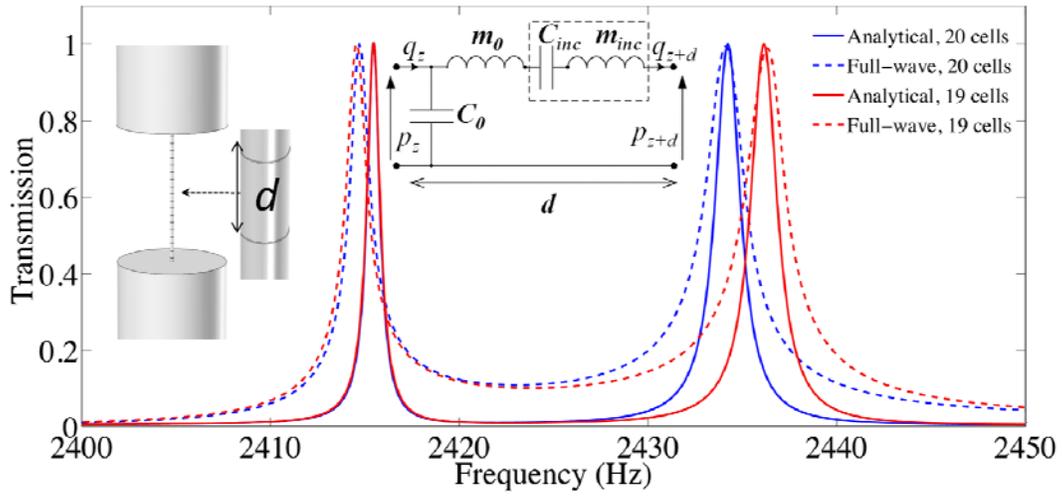

Figure 1 – Energy transmission coefficient versus frequency for two ultranarrow channels with different lengths. The left peak, independent of the channel length, corresponds to DNZ operation. The insets show the geometry under analysis and the acoustic circuit equivalent of a unit cell of length $d$ containing a membrane ($m_{inc} // C_{inc}$ series element).

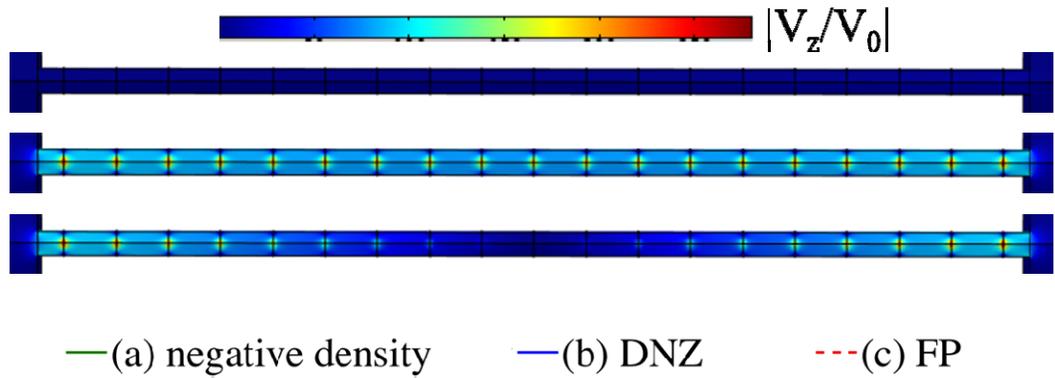

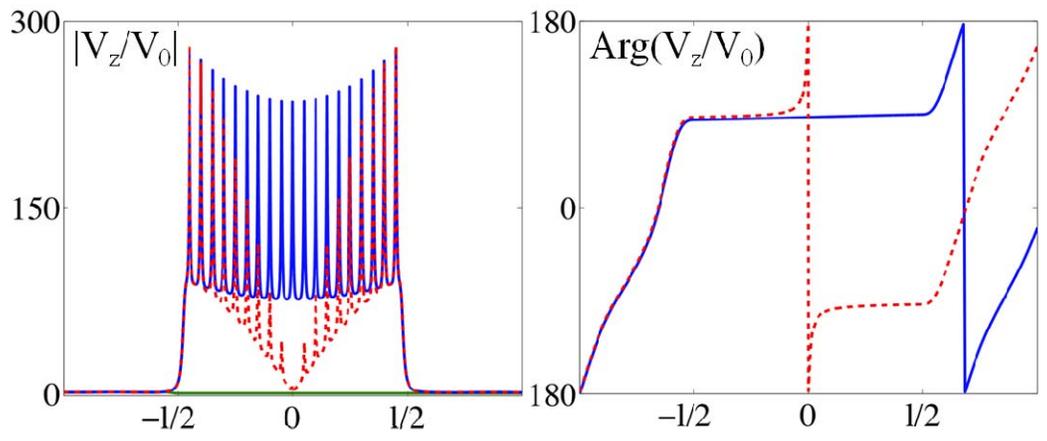

Figure 2 –Top: Velocity field distribution (magnitude) along the channel operating at (a) negative density, (b) DNZ condition and (c) first FP resonance. Bottom left: Same, but along the center line of the channel. Bottom right: Phase of the particle velocity along the same line.

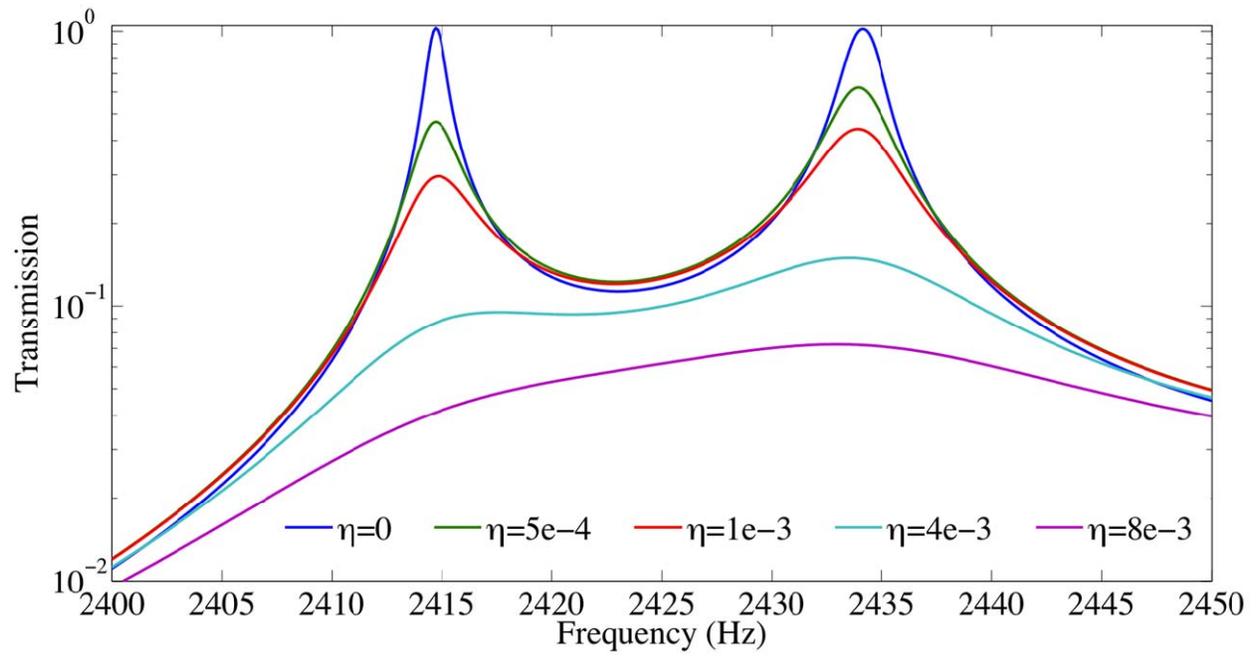

Figure 3 – Effect of membrane losses on the energy transmission coefficient for a 20 cell channel around the DNZ frequency.

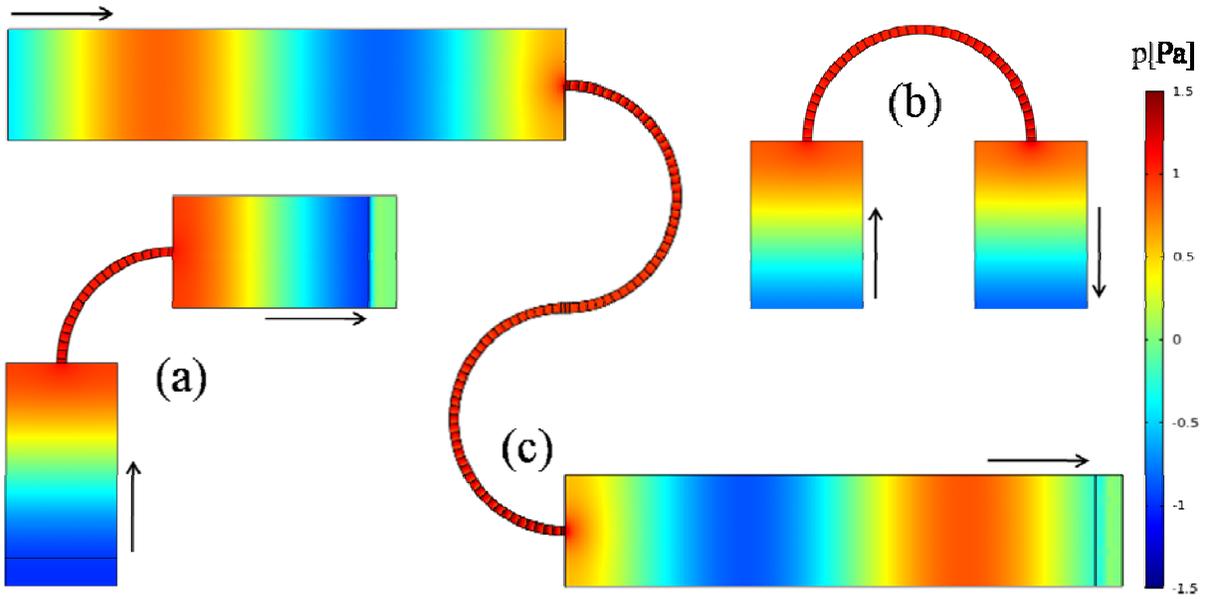

Figure 4 – Acoustic pressure field distribution (snapshot in time) in waveguides coupled through (a) a 90° bent channel, (b) a 180° bent channel and (c) a S shaped channel, operated at their common DNZ frequency. The arrows represent the direction of the wave vector.